# Influences of the dissipative topological edge state on quantized transport in MnBi$_2$Te$_4$

Weiyan Lin[1,†], Yang Feng[2,8,†], Yongchao Wang[4,†], Zichen Lian[3], Hao Li[5,6], Yang Wu[6,7], Chang Liu[3,8], Yihua Wang[2,9], Jinsong Zhang[3,10], Yayu Wang[3,10], Xiaodong Zhou[1,12,13*] and Jian Shen[1,2,9,11,12,13*]

[1]State Key Laboratory of Surface Physics and Institute for Nanoelectronic Devices and Quantum Computing, Fudan University, Shanghai, China.

[2]Department of Physics, Fudan University, Shanghai, China.

[3]State Key Laboratory of Low Dimensional Quantum Physics, Department of Physics, Tsinghua University, Beijing, China.

[4]Beijing Innovation Center for Future Chips, Tsinghua University, Beijing, China.

[5]School of Materials Science and Engineering, Tsinghua University, Beijing, China

[6]Tsinghua-Foxconn Nanotechnology Research Center, Department of Physics, Tsinghua University, Beijing, China.

[7]Department of Mechanical Engineering, Tsinghua University, Beijing, China.

[8]Beijing Academy of Quantum Information Science, Beijing, China

[9]Shanghai Research Center for Quantum Sciences, Shanghai, China

[10]Frontier Science Center for Quantum Information, Beijing, China.

[11]Collaborative Innovation Center of Advanced Microstructures, Nanjing, China

[12]Shanghai Qi Zhi Institute, Shanghai, China

[13]Zhangjiang Fudan International Innovation Center, Fudan University, Shanghai, China

[†]These authors contributed equally to this work
*Emails: zhouxd@fudan.edu.cn, shenj5494@fudan.edu.cn

Abstract:

The beauty of quantum Hall (QH) effect is the metrological precision of Hall resistance quantization that originates from the topological edge states. Understanding the factors that lead to quantization breakdown not only provides important insights on the nature of the topological protection of these edge states, but is beneficial for device applications involving such quantized transport. In this work, we combine conventional transport and real space conductivity mapping to investigate whether the quantization breakdown is tied to the disappearance of edge state in the hotly studied MnBi$_2$Te$_4$ system. Our experimental results unambiguously show that topological edge state does exist when quantization breakdown occurs. Such edge state is dissipative in nature and could lead to a quantization breakdown due to its diffusive character causing overlapping with bulk and other edge states in real devices. Our findings bring attentions to issues that are generally inaccessible in the transport study of QH, but can play important roles in practical measurements and device applications.

## I. INTRODUCTION

Similar to quantum Hall (QH) effect[1], quantum anomalous Hall (QAH) effect[2,3] originates from the topologically protected chiral edge state. The QAH phase was firstly established in magnetically doped topological insulator (TI) although its onset temperature remains significantly lower than the Curie temperature despite of tremendous efforts to raise it[4-7]. While such quantization breakdown, i.e., deviation from $h/e^2$ in $R_{yx}$, is generally attributed to the dissipation added to the chiral edge conduction[8], much remains unknown regarding the source of such dissipations. Early non-local transport has suggested the existence of dissipative nonchiral edge mode[9,10]. However, this edge-dominated dissipation picture was later challenged by transport using Corbino geometry which identified bulk conduction as the dominant source of dissipation[11]. Another source of dissipation comes from the magnetic disorder leading to strong spatial fluctuations of exchange energy gap[12,13]. An essential question is whether the chiral edge state could survive with such disorders. This issue becomes relevant for intrinsic magnetic TI MnBi$_2$Te$_4$[14-17] where angle resolved photoemission spectroscopy fails to detect the exchange gap in its surface Dirac cone below Neel temperature[18-20] casting doubts on its supremacy in terms of the magnetic ordering. Moreover, a multi-domain state could form inside the material resulting in a complex network of conducting edge channel that would affect the quantized transport as well[21,22].

In this work, we investigate the underlying mechanism of quantization breakdown in MnBi$_2$Te$_4$ which can be viewed as layered TI Bi$_2$Te$_3$ with each of its Te-Bi-Te-Bi-Te quintuple layers intersected by a Mn-Te bilayer, forming Te-Bi-Te-Mn-Te-Bi-Te septuple layer (SL). The magnetic moment of Mn orders ferromagnetically (FM) within each SL, and antiferromagnetically (AFM) between neighboring SL. When all the Mn moments are aligned parallel by an external magnetic field, MnBi$_2$Te$_4$ is driven to a Chern insulator with quantized Hall resistance. The zero-field QAH has been observed on 5-SL MnBi$_2$Te$_4$ at 1.4 K[23]. Under external field, this quantization temperature can go up to a few tens of kelvin, an encouraging sign for potential applications[24,25]. The AFM ground state of MnBi$_2$Te$_4$ is also a topological phase different from the Chern insulator. Previous transport measurement has identified a zero-Hall plateau (ZHP) on 6-SL MnBi$_2$Te$_4$ in its AFM state suggesting an axion insulator[26]. More interestingly, a dissipative edge state was recently reported in such a ZHP phase by non-local transport[27] and scanning microwave impedance microscopy (sMIM) studies[28]. Different from previous works, we complement conventional transport with real space conductivity mapping to acquire local conducting properties essential for interpreting global transport behavior. By observing how the quantized transport is "destroyed" by gating and elevating temperature, we identify the dissipative edge state as the origin of the observed quantization breakdown, rather than the multi-domain state or the absence of the edge conduction channel. We also discuss the influence of such dissipative channel on the edge conduction in real devices giving insights to the observed transport behavior.

## II. RESULTS AND DISCUSSION

We first look at the gate voltage dependent transport and sMIM characterization of the Chern-insulator phase. sMIM measures the local conductivity with nanoscale spatial resolution which can distinguish an insulating bulk from a conducting edge in topological systems[29]. Figure 1(a) is an optical image of one MnBi$_2$Te$_4$ sample (sample I), on which the transport and sMIM data were measured from the 6-SL thickness area (purple color). Figure 1(b) shows the gate voltage dependent longitudinal resistance $R_{xx}$ and Hall resistance $R_{yx}$ measured at 1.7 K and 9 T. As expected, $R_{yx}$ gradually approaches to the quantized value $h/e^2$ with increasing gate voltage while $R_{xx}$ vanishes at the same time. What is unusual is that $R_{xx}$ seems to be quantized after $V_g = +74V$ while $R_{yx}$ continues to rise from $V_g = +74V$ to $V_g = +80V$. This behavior is in contradiction to conventional wisdom that quantization of $R_{xx}$ and $R_{yx}$ should happen simultaneously. To understand such transport behavior, we use sMIM to probe the gate evolution of conducting properties at microscopic scale. Figure 1(c) shows a series of gate dependent sMIM images taken in the area A marked in Fig. 1(a). Here larger sMIM signal corresponds to higher local conductivity. The sample's interior (bulk) changes from a metallic to an insulating state as its chemical potential moves into the gap by gating. At the same time, a persistent metallic state exists at the sample's edge corresponding to the in-gap topological edge state. What's more, the sample edge experiences a stronger electric field and thus a larger effective doping. As shown in Fig. 1(c), the insulating area of the bulk first appears near the edges and then spreads inwards. An intriguing situation happens at $V_g = +70V$ where a conductive bulk and a metallic edge are spatially separated by an insulating strip in between. This situation becomes more apparent at $V_g = +74V$. This inhomogeneous gating originates from a geometric effect[30] and is of particular relevance for our transport measurement as discussed below. In Fig. 1(d), we extract the bulk and edge sMIM signals and plot them as a function of the gate voltage. These sMIM signals can be converted to real conductivity (see supplementary information S1). In a gating experiment, the Hall quantization breakdown, i.e., deviation from $h/e^2$ in $R_{yx}$, is usually attributed to the bulk conduction. If one compares $R_{yx}$ in Fig. 1(b) to sMIM-bulk in Fig. 1(d), they indeed show a very similar gate voltage dependence indicating their close relationship. We examine this relationship by performing a model simulation to calculate $R_{yx}$ and $R_{xx}$ using the measured bulk and edge conductivity (see supplementary information S2). This model adopts Landauer-Buttiker formalism to describe the ballistic transport of chiral edge channels. The conducting bulk is treated as a parallel channel with the edge modifying the transmission probability between two terminals as done before[8]. In Fig. 1(b), the simulated results lay nicely over the transport curves except for $R_{yx}$ at higher gate voltages. In particular, this model reproduces the $V_g = +74V$ case where $R_{xx}$ is quantized to 0 while $R_{yx}$ is only 0.75 $h/e^2$. This can be understood in the following picture. At $V_g = +74V$, the device possesses two spatially separated current flow channels. Roughly 75% of injected current goes along the dissipation-less edge while the remaining 25% goes through the bulk. Since there is no cross-talk between the edge and bulk channel, the edge channel remains dissipation-less and thus already "quantized" giving vanishingly small $R_{xx}$ and contributing to 0.75 $h/e^2$ in $R_{yx}$. As the bulk becomes more insulating at higher gate voltages, more current will flow along the edge and the "quantized" value of $R_{yx}$ will increases accordingly. Our model predicts a 100% quantization of $R_{yx}$ after $V_g = +77V$ based on the measured bulk conductivity while the experiment doesn't reach this level indicating residual bulk current[31,32]. Nevertheless, our sMIM imaging and model simulation corroborate the believe that the bulk conduction is a major source of Hall quantization breakdown in a gating experiment. More importantly, it highlights the importance of knowing the actual current flow in real devices to understand the seemingly counterintuitive transport results.

We next investigate the temperature dependence of the Chern insulator phase, which is mostly relevant to quantization breakdown. The fact that QAH temperature of MnBi$_2$Te$_4$ is still much lower than its Curie temperature falls short of the expectation for this stoichiometric compound and leaves open various possibilities to account for such discrepancy. Here we use sMIM imaging to narrow down the possible scenarios. Figure 2(a) and (b) shows $R_{xx}$ and $R_{yx}$ versus temperature and magnetic field on the same device shown in Fig. 1. At 1.7 K, this 6-SL MnBi$_2$Te$_4$ displays a magnetic field driven topological phase transition from a ZHP phase with C=0 to a Chern insulator phase with C=1. A clear plateau-to-plateau transition happens in $R_{yx}(H)$. When the temperature increases, this plateau-to-plateau transition becomes broad signaling the quantization breakdown. Following previous attempts of QH analysis[33], we plot $R_{xx}(9T)$ as a function of temperature in a log-log scale in the upper panel of Fig. 2(c). The data points fall naturally on a straight line indicating a power law dependence on temperature. Previous temperature dependent QH study has seen a change from a power law to an exponential dependence of $R_{xx}(T)$ as the temperature increases[33], corresponding to a crossover from variable range hopping to thermal activation behavior of charge transport. The fact that we only see a power law up to 30 K implies that thermal activation is still "frozen out" in this temperature range. In the bottom panel of Fig. 2(c), we plot $\Delta R_{yx}$, which is the deviation of $R_{yx}$ from $h/e^2$, and $R_{xx}$ together. A linear relationship has been observed similar to QH systems before[34,35]. This empirical result reconfirms the consensus that quantization breakdown in QH system is due to the dissipation. What remains to be answered is the source of such dissipation.

Our sMIM imaging provides clue to origins of dissipation. Figure 2(d) shows the sMIM images of the Chern insulator phase at various temperatures. They look almost identical with an insulating bulk interior surrounded by a conductive edge. The only change perceived from these images is the widening edge state with increasing temperatures. Such apparent width in sMIM images cannot be taken as a real physical dimension of topological edge state, but rather a measure of how metallic edge states penetrate into the insulating bulk. We apply a Gaussian fit to extract such penetration depth of edge states which increases from 2 to 3 μm in this temperature range (see supplementary information S3). In regarding the dissipation source, the sMIM observation has the following implications. First, this quantization breakdown is not caused by the absence of chiral edge state which continues to exist even $R_{yx}$ is far below $h/e^2$ at high temperatures. Similar conclusion has been drawn in magnetically doped TI system[36]. Second, there is no internal current flow due to the formation of multi-domain states. Third, bulk conduction should be insignificant as sMIM imaging shows that bulk remains insulating without thermal activation. Last but not least, we can only identify a widening edge state as an experimental signature which strongly suggests an edge-related dissipation mechanism leading to a diffusive edge channel. For instance, a recent theoretical work considers the thermal spin fluctuations as the origin of quantization breakdown at high temperatures[37]. Such thermal spin fluctuations act as frozen magnetic disorders that scatter the chiral edge state and facilitate its overlapping with the bulk states, which may be the origin of quantization breakdown.

The gating and temperature dependent experiments above illustrates the need of preserving a dissipation-less nature of edge state in achieving a quantized transport in MnBi$_2$Te$_4$. We now show another experiment to make this point clearer. Figure 3(a) is the optical image of another 6-SL MnBi$_2$Te$_4$ (sample II) with multiple pairs of electrodes for transport characterizations. Different from sample I, two cracks exist in the interior of this device denoted by orange arrows in Fig. 3(a). There are also two notches at the fringe indicated by red arrows. Figure 3(b) is a schematic of the transport setup with all the electrodes properly marked. We run the current from source K to drain F and measure Hall resistance $R_{yx}$ from a pair of opposite-placed electrodes perpendicular to the current flow. In Fig. 3(c), we plot $R_{yx}(H)$ measured from 4 pairs of electrodes (GE, HC, IB and JA). While the overall shape of $R_{yx}(H)$ are similar with a clear plateau-to-plateau transition, there exists a systematic voltage offset for ZHP among those electrodes. For electrodes that are further from the cracks such as GE, we obtain the correct $R_{yx}(H)$ as reported before[26]. For electrodes that sit near the cracks (HC, IB, and JA), there is an offset for ZHP. Interestingly, such offset only happens for ZHP, but not for the Chern insulator phase, i.e., $R_{yx}(H)$ from 4 pairs of electrodes all collapse to $h/e^2$ plateau at high fields. We note that $R_{yx}(H)$ in Fig. 3(c) are raw data without anti-symmetrization as routinely applied in Hall measurement. Such anti-symmetrization in $R_{yx}$ is intended to remove components symmetric with magnetic field, i.e., $R_{xx}$ components. For example, the misalignment of opposite-placed Hall electrodes along current flow direction usually gives rise to such artifacts which should be removed by anti-symmetrization. However, the systematic ZHP offset we observe here cannot be attributed to this problem as seen in the optical image. Instead, it reflects the property pertain to this device which we will discuss. Fig. 3(d) shows sMIM images acquired from the crack area, which are taken at zero and 9 T corresponding to ZHP and Chern insulator phase, respectively. The most important feature revealed in such imaging is that, the dissipative edge state of ZHP phase is spatially more extended than the dissipation-less chiral edge state of Chern insulator phase. In addition, we see a conductive strip along the crack in Chern insulator phase corresponding to the topological edge state. The same edge state should exist in ZHP, but is hard to be resolved due to the spreading edge state nearby. This difference in terms of the dissipation level of edge states has a dramatic consequence on the charge transport in the device.

For a better understanding, let us start with $R_{xx}$ shown in Fig. 4(a) as a function of magnetic field for 7 pairs of electrodes. They all vanish at high fields signaling a dissipation-less edge state in the Chern insulator phase. However, $R_{xx}$ acquires a finite value at ZHP phase due to the dissipative nature of the edge channel. Figure 4(b) shows the voltage drop $V$ at zero field of each pair of electrodes as a function of the longitudinal distance $L$ between the electrodes. We expect these data points (diamonds in Fig. 4(b)) should converge onto a straight line whose intercept is 0 dictated by the Ohm's law for such a dissipative edge. In fact, only three points (AB, DE and HG) fall on the line with the relation $V = 0.75 \times L$. The voltage drops of JI and BC, indicated by the green and purple diamonds are not far from the line. Interestingly, if we replace the distance between the electrodes JI and BC by the length of the notch, i.e., black solid curve in Fig. 4(c) connecting neighboring electrodes, we have the new data points of JI and BC (green and purple circles) be located exactly on the straight line obeying Ohm's law. However, two remaining data points IH and CD cannot fit into this line even considering the replacement of the distance by the actual boundary length which shows tiny difference without the notch shape (the same for other pairs of electrodes). IH and CD display a larger resistance than expected, which is likely caused by the interaction between the crack and the edge in these areas as denoted in Fig. 4(c). The diffusive character of edge state at zero field facilitates the overlapping between the crack and the edge ultimately causing a larger dissipation level. In Fig. 4(d), we plot the relative voltage drop among those electrodes according to our analysis above which naturally explains the voltage offset for ZHP in Fig. 3(c). Our previous work suggests a quantum spin Hall origin of the edge state at zero field in 6-SL MnBi$_2$Te$_4$[28]. Such helical edge state is less immune to the impurity scatterings which makes the quantized transport fragile[38]. On the other hand, for Chern insulator phase, the chirality of edge state makes it robust against backscattering and preserve its dissipation-less nature explaining the nice quantization of $R_{xx}$ and $R_{yx}$ among those electrodes.

## III. CONCLUSION

We discuss the factors that may lead to quantization breakdown of topological edge conduction in MnBi$_2$Te$_4$. While the topological edge state can survive under different circumstances in the absence of quantization, preserving its dissipation-less nature is the key to achieve quantized transport. One consequence of such dissipative edge state on quantized transport is to cause spatial overlapping between edge and the bulk state resulting in unexpected transport behavior. Moreover, this work demonstrates sMIM as an important probe to complement the conventional transport. Knowing the microscopic details of the local current distribution is essential to comprehensively understand the behavior of global charge transport.

**Acknowledgements**

We acknowledge the discussions with C. Z. Chen and Y. H. Li. The work at Fudan University is supported by National Natural Science Foundation of China (Grant Nos. 12074080, 11804052, 11904053 and 11827805), National Postdoctoral Program for Innovative Talents (Grant No. BX20180079), Shanghai Science and Technology Committee Rising-Star Program (19QA1401000), Major Project (Grant No. 2019SHZDZX01) and Ministry of Science and Technology of China (Grant Nos. 2016YFA0301002 and 2017YFA0303000). The work at Tsinghua University is supported by National Natural Science Foundation of China (Grant Nos. 21975140, 51991313), the Basic Science Center Project of National Natural Science Foundation of China (Grand No. 51788104) and the National Key R&D Program of China (Grand No. 2018YFA0307100).

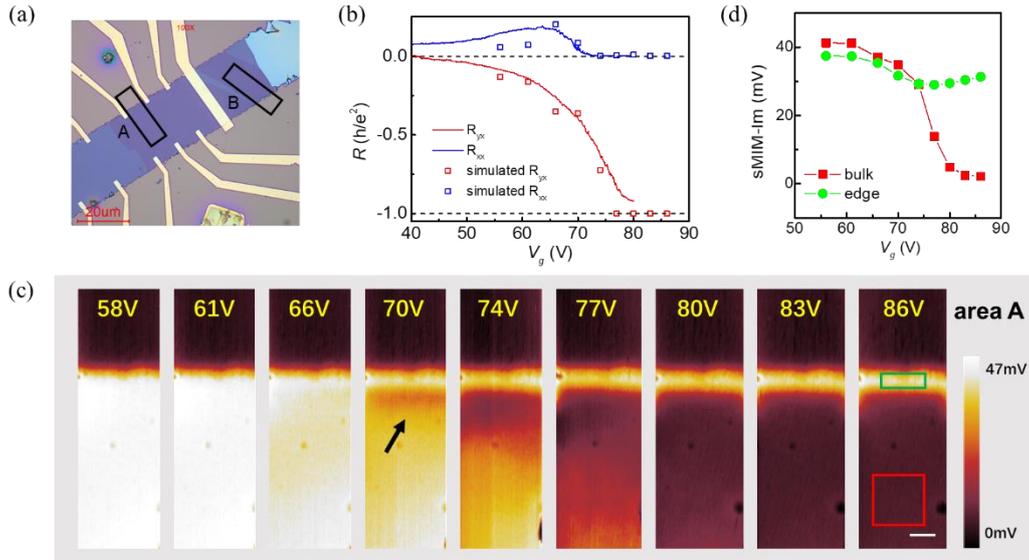

FIG. 1. (a) Optical image of sample I. (b) Gate voltage dependent longitudinal resistance $R_{xx}$ and Hall resistance $R_{yx}$ at 9 T and the Landauer-Buttiker model simulation results. (c) Gate voltage dependent sMIM images at 9 T of area A marked in (a). Area with the electric field focusing is denoted by the black arrow. Scale bar is $2\mu m$. (d) Gate voltage dependent sMIM bulk (edge) signal averaged from the red (green) rectangular area denoted in (c).

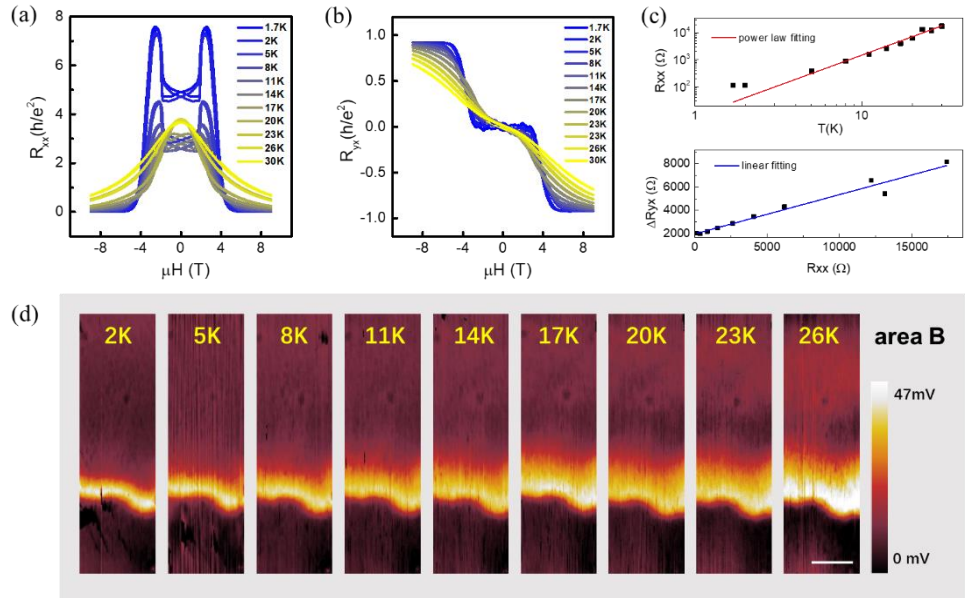

FIG. 2. (a)(b) Longitudinal $R_{xx}$ and Hall resistance $R_{yx}$ versus magnetic field acquired at various temperatures. (c) Power law fitting of $R_{xx}(T)$ at 9 T and the linear relationship between $R_{xx}$ and $\Delta R_{yx}$ in the Chern insulator phase. (d) Temperature dependent sMIM images at 9 T of area B marked in Fig. 1(a). Scale bar is $4\mu m$.

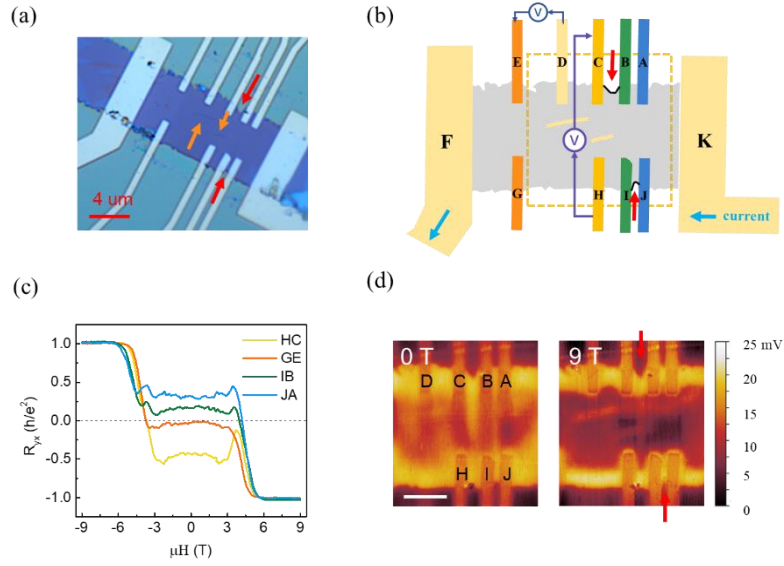

FIG. 3. (a) Optical image of sample II. (b) Schematic of transport measurement setup. sMIM scan field of view is marked by dash orange square. (c) Hall resistance $R_{yx}$ as a function of magnetic field for 4 pairs of electrodes. (d) sMIM images of crack area taken at 0 and 9 T. Scale bar is $4\mu m$. Red arrows in (a), (b), and (d) indicate notches near electrodes BC and JI at the fringe of the device. Orange arrows in (a) indicate cracks inside the device.

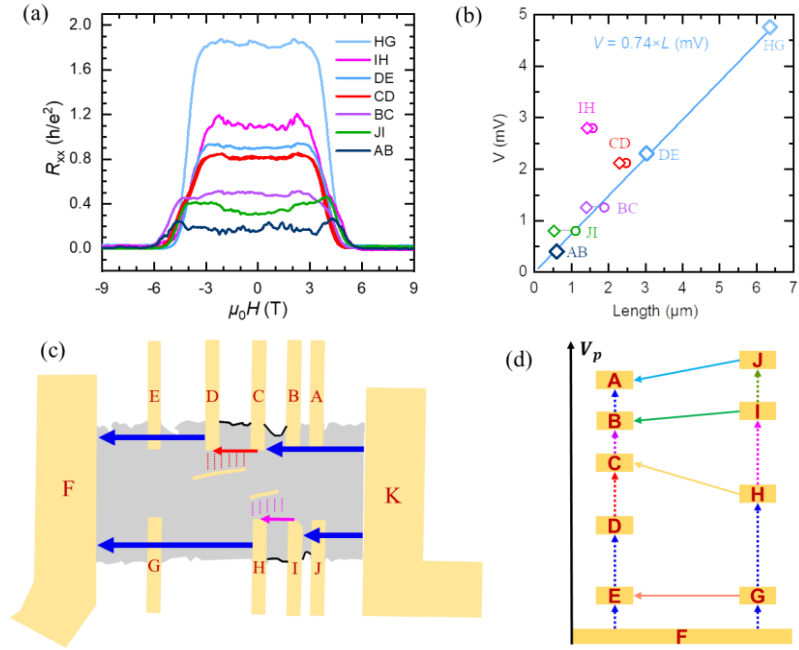

FIG. 4. (a) Longitudinal resistance $R_{xx}$ as a function of magnetic field for 7 pairs of electrodes. (b) Voltage drop of various pairs of electrodes as a function of the distance between two electrodes in a pair (diamonds) or the length of sample boundary connecting neighboring electrodes (circles). (c) Schematic of current flow pattern in the device at zero field. Red shadow and pink shadow represent the interaction between sample boundaries and internal cracks. (d) Schematic of relative potential among various electrodes.